# The latent potential of *YouTube* – Will it become the 21$^{st}$ Century Lecturer's Film Archive?

**Adam Micolich\***
School of Physics
The University of New South Wales
Sydney NSW 2052
Australia

mico@phys.unsw.edu.au

Abstract: YouTube *(http://www.youtube.com) is an online, public-access video-sharing site that allows users to post short streaming-video submissions for open viewing.  Along with* Google, MySpace, Facebook*, etc. it is one of the great success stories of the Internet, and is widely used by many of today's undergraduate students.  The higher education sector has recently realised the potential of YouTube for presenting teaching resources/material to students, and publicising research.  This article considers another potential use for online video archiving web sites such as* YouTube *and* GoogleVideo *in higher education – as an online video archive providing thousands of hours of video footage for use in lectures.  In this article I will discuss why this might be useful, present some examples that demonstrate the potential for* YouTube *as a teaching resource, and highlight some of the copyright and legal issues that currently impact on the effective use of new online video web sites, such as* YouTube*, for use as a teaching resource.*

## Introduction: What is *YouTube* and why is it useful?

*YouTube* is a public-access online video-sharing site where users can post short video clips (up to 10 minutes in length) for open viewing.  As such, it is a member of a family of web sites including the social networking sites *MySpace* and *Facebook*, image servers such as *Photobucket* and *Flickr*, online journal sites (known as 'blogs') such as *Blogspot*, and online encyclopedias such as *Wikipedia*, that have evolved to cater to the public's desire to share information and express themselves as individuals via the Internet.

In its original incarnation, *YouTube* mostly contained short video snippets that users filmed with digital video cameras and posted on *YouTube*'s web site so that their friends could view them online.  However, in the last couple of years, *YouTube* has grown to be so much more as businesses, community groups and political organisations have realised *YouTube*'s potential for advertising products, presenting policies, etc. to the public.  Most recently, the higher education sector has even taken notice, with universities beginning to open their own *YouTube* channels to publicise research, present teaching resources in a more accessible way, and attract new students (Alexander 2007; Anwar 2007).  However, there is a dimension of *YouTube* applicable to teaching that to my knowledge has not been identified in all this – the potential use of *YouTube* as a digital video library for teaching purposes.

As the public has realised the full potential of video sharing web sites, and technologies such as digital video cameras and hard-disk video recorders have become commonplace, the quantity, quality and diversity of video postings on *YouTube* and similar sites has increased rapidly.  Now, almost anything can be found on *YouTube*, which currently hosts over 83 million videos and receives over 50,000 new contributions per day (Wikipedia 2008a): interesting news clips, interviews with public figures, footage of interesting physical phenomena, tutorials on how to do various things, computer generated simulations, time-lapse videos, etc.  In a sense, *YouTube* has now evolved from an amateur film-sharing site into an enormous, online film archive with easy access from your computer, 24 hours a day, 7 days a week.  Furthermore with *YouTube* as inspiration, numerous other online video web sites have now emerged (e.g., *GoogleVideo, DailyMotion, Brightcove, iFilm,* etc.), rapidly increasing the quantity of video content available online.  Nothing like this has ever existed before, and as a lecturer seeking short snippets of video footage to present in class, *YouTube* and similar online video archive web sites are a dream come true.

**Why might one use *YouTube* videos in physics lectures?**

Physics is a challenging subject that combines experimental observations, mathematical descriptions, and which calls upon physical intuition to provide scientifically tested explanations for the various physical phenomena. The first two aspects are comparatively easy to teach and relatively well catered to by traditional laboratory classes, lectures, and tutorials. The intuitive aspect of physics, however, is more nebulous and difficult to convey, and usually comes to a student indirectly via accumulated experience (Kolb 1984), for example, by noting that common skills and techniques can be applied to often quite different physical systems.

One way to facilitate the development of physical intuition and accelerate the process of gaining 'experience' is to provide a strong background context for the technical aspects of the course, either by performing live demonstrations or by presenting relevant examples and anecdotes in the lectures. Although the efficacy of live demonstrations is established (Di Stefano 1996), they tend to become less viable in higher year courses because more sophisticated, expensive and cumbersome equipment is required as the physics becomes more complex. For example, while demonstrations of basic physics can be done easily using simple apparatus such as masses, springs, batteries, beakers of water, etc., the same is not true for quantum physics, where the effects can often only be observed at the atomic scale. Additionally, the set of useful demonstrations is often smaller in higher year courses because the students have often already seen many of the most common and best ones in their first year physics courses. Hence, the onus tends to fall back on examples and anecdotes or worked problems in higher year courses. However, in my experience as both a student and as a lecturer, examples and anecdotes are often not as effective because they lack the visual dimension and dynamic of live demonstrations.

One way to enhance the value of examples and anecdotes is to use video footage. Some reasons why you might want to do this include:
a) supplementing the limited number of viable live physics demonstrations available by using video footage as *de facto* demonstrations of physics that could never otherwise be done in a lecture theatre (e.g., detonating a nuclear weapon or freezing a lake);
b) providing a visual dimension to many of the examples and anecdotes given in the lecture to enhance their effectiveness and memorability; and
c) making the course more interesting, engaging and more relevant to students' everyday experiences.

In the past, the use of video in lectures was rather difficult unless one was in the fortunate and expensive position of having access to a very large film archive or was willing to invest the considerable time and effort involved in producing his or her own video clips. The advent of *YouTube*, digital video processing software and digital presentation technologies in lecture theatres has the potential to change all this – in the future a lecturer may have online access to thousands of hours of video footage, that is more easily searched, prepared and presented than ever before.

It should be noted that although I discuss the potential of using *YouTube* as a resource for teaching physics in this article, these ideas are by no means limited to this subject alone, or to the sciences in general. In fact, I find it hard to think of a subject that could not benefit from this approach, for example, a politics course could use news clippings or media interviews with politicians, or a mechanical engineering course could take a guided tour through an automotive factory in another country to see the latest in robotics technologies – the possibilities are almost endless.

In the following sections, I will discuss the potential for using *YouTube* and other similar online video web sites as a film archive for use in lectures. I will begin with a discussion of the copyright and legal aspects of using online video in lectures. Not only are these requirements quite limiting in terms of the potential for online video in lectures, but using the web interface in the lecture environment can also be problematic. Given this, I will then, legal considerations aside, provide a comparative discussion of some of the possible routes to implementing online video as a teaching resource. This will be followed with a discussion of the potential motivations for using online video in lectures, including some examples of content that demonstrate these motivations and the potential benefits of using online video in lectures. I will conclude by highlighting the need for new copyright structures that, in conjunction with providing fair protection to the creators of video content, also allow the education sector to capture the potential benefits that online video archive web sites such as *YouTube* have to offer.

**Copyright and legal aspects of using online video in lectures**

There are two important legal aspects that need to be considered in contemplating the potential for using online video in lectures. The first is **Copyright** and the second is the legal **Terms of Use** for the various web sites that might be used to source those videos.

**Copyright**
Copyright law aims to provide a balance between rewarding the creators of intellectual material for their work, while at the same time allowing reasonable and useful access to that material by consumers. In a sense, it is an agreement between the public and the creators of intellectual material, providing certain limited exclusive rights that ensure a fair marketplace for creators to distribute their content and reap the rewards that act as an incentive to produce the content in the first place. From here onward I will discuss copyright from an Australian context. While the basic principles are somewhat common, the exact details of copyright law vary from country to country, and you should check your local legislation where relevant.

In Australia, the Part VA and VB statutory licences of the Copyright Act (Copyright Act 1968) allow the use of copyrighted material for educational purposes without the need to obtain direct permission from the copyright owner. While Part VA covers the copying and use of audiovisual and broadcast material, most web-based media (i.e., webcasts) does not fall under the definition of broadcasting as outlined under Part VA of the Act. Instead, most online video falls under Part VB, which is somewhat more limiting regarding what can be done with the copied material. The requirements under the Part VB licence are:

a) the material must be restricted to only staff and students of the educational institution, and all reasonable steps must be taken to ensure that material is not accessible to the general public;
b) no more than 10% of a work can be used or made available online at any one time; and
c) an 'electronic use notice' indicating that the material has been copied under licence must be placed so that the material cannot be reached without the notice being viewed.

It should be noted that Parts VA/VB only apply when the work is that of the original owner – if someone has made an illegal copy, and you use or copy that illegal copy, then that is an infringement of copyright. This is an important issue when it comes to online video. Submissions to *YouTube* and other online video web sites can be broadly classified into three groups (O'Brien and Fitzgerald 2006). The first is original creations that have been posted to the web site by their creator and legal copyright holder, for example, home videos, original short movies and films, etc. This material is quite amenable to use providing that the owner gives permission to use it. Contacting the owner is usually rather simple, and the various video web sites generally facilitate direct contact with users who post video content to the web site.

The second group consists of 'transformative derivatives' – film that consists of some mix or remix of original content altered to such an extent that it is something new and creative. And the third group consists of copied material that has been posted to the web site by someone other than its creator, who may or may not have the permission of the legal copyright holder to post that material to the web site. Considerable care needs to be taken with material in the latter two groups, because a large proportion of these videos infringe copyright, ranging from blatant infringement through to material that on the surface appears quite benign in terms of 'fair use' but which could be competently argued in court as copyright infringing content. Note that although web sites such as *YouTube* have a moral obligation to prevent infringing material from appearing on their web site, despite their best attempts such content still appears, largely because it is somewhat impossible to check approximately 50,000 videos that are posted each day for infringing content.

If usage beyond that covered by Parts VA/VB is required (e.g., you exceed the licence limits, or edited versions or derivative works need to be produced from the content), then the direct permission of the legal copyright holder for the material must be obtained instead.

**Terms of use**

Over and above copyright law are the terms of use of any web sites that you might use to source video material. These govern your legal use of the web site, and commonly take the form of an implied contract between you and the web site owner that you enter into automatically simply by using the web site. The terms of use vary between web sites, and some are more limiting than others. For example, *YouTube* is rather restrictive regarding the use of video content hosted on its web site, in particular, they require that (YouTube 2008):

a) "4C. A user agrees not to access user submissions or *YouTube* content through any technology or means other than the video playback pages of the website itself, the *YouTube* embeddable player…"; and
b) "5A. …Content on the website is provided to you *as is* for your information and personal use only and may not be downloaded, copied, reproduced, distributed, transmitted, broadcast, displayed, sold, licensed, or otherwise exploited for any other purposes whatsoever without the prior written consent of the respective owners…".

In contrast, *Google video* is slightly less restrictive, specifying instead that (GoogleVideo 2008):

a) "8.1. You understand that all information (such as data files, written text, computer software, music, audio files or other sounds, photographs, videos or other images) which you may have access to as part of, or through your use of, the [*Google*'s] services are the sole responsibility of the person from which such content originated…"; and
b) "8.2. …You may not modify, rent, lease, loan, sell, distribute or create derivative works based on this Content (either in whole or in part) unless you have been specifically told that you may do so by *Google* or by the owners of that Content, in a separate agreement."

Note that copyright and terms of service are not entirely complimentary, such that satisfying the requirements of one will not necessarily satisfy the requirements of the other. I will now look at some various possibilities for using online video in lectures. It should be noted that the legality of these approaches will depend on the content you intend to use, whether or not you have obtained permission to use it, and the terms of use of web site you might be sourcing that content from. Where appropriate below, I will comment further on the legal aspects, but in all cases you should check that you meet the relevant copyright legislation and terms of use for each video/web site that you use.

**The possibilities for using online videos in lectures**

There are a number of possible routes to utilising online video as an educational resource. In this section I will explore three of the most obvious possibilities, with the latter being what I see as the most ideal possible route in terms of ease, effectiveness and reliability.

**1. Providing hyperlinks for use after the lecture**
By far the easiest option for utilising online video as an educational tool is to simply provide a hyperlink (URL), so that the students can watch the desired video on their own time after the class. This option is also the least effective. Firstly you need to rely on the students bothering to watch the video after the class, they will not always do this, so the video is wasted for some portion of the students (in my experience this can be 95% or more). For those who do view the video, it won't necessarily be clear to them what parts of the video are important, particularly in the case of longer videos where you are only interested in some small portion of the content. Additionally, there is also some risk that the students will misinterpret the connection between the content of the video and the concepts that you intend it to demonstrate.

These problems result from a fundamental disconnect between viewing and discussion, which is a serious limitation of this approach, almost to the point where the video becomes little more than trivia unless the video is well chosen such that its content-concept relationship is obvious. Unfortunately, such videos are a very small subset of the multitude of useful possible videos that exist.

It is worth noting one route to improving this approach, which is to use the 'embed' tag on *YouTube* to incorporate videos into teaching web sites, discussion forums and lecturer blogs. This allows a lecturer to add commentary text and notes around the video that aid its use by the students. However, despite being very useful, this approach is often not as powerful as presenting the video in a classroom environment.

**2. Showing the video using a web-interface**
Another possibility would be to display the video during the lecture using the web-based video player embedded in the web site. This would be relatively straightforward to implement – most lecture theatres now have internet connections to the computers that control the display projector, and it is very easy to put an internet hyperlink into a *PowerPoint* presentation.

Compared to giving the students a hyperlink to view later, this approach has a distinct advantage, namely that it is possible to introduce the video, present it to the class, and then discuss its content afterwards. Not only does this allows a lecturer to more effectively link the content of the video to the concepts being taught, but it provides an opportunity for a whole additional layer of in-class discussion that would never occur if the students view the same footage individually on their own time. In fact, when the students view footage outside class, it is almost never discussed again – either between the lecturer and the students, or the students themselves. However, there are still a number of problems attached to this approach.

- First and foremost, the Internet is not 100% reliable, everything from the web site itself through to the lecture theatre's web connection will fail at some point. So occasionally you will try to connect to the video on the web site and nothing will come up. While students often find this funny, it can be highly distracting;

- Even if the Internet is working, it can sometimes be slow. Hence a video that plays perfectly well when viewed in your office may stop and start, stall entirely, or take a long time to load when it is played in the lecture because the streaming video data is arriving too slowly;

- Although the embedded video players in most web sites such as *YouTube* are of good quality, they usually only take up a small fraction of the browser window, which means that the video can be difficult to see clearly in the lecture theatre (although you can run the player in full-screen mode, the resolution can sometimes be poor). Additionally, other items on the web site around the video player (e.g., comments, advertising, etc.) can be highly distracting in a lecture environment; and

- Finally, you have to take the stream as it comes. For example, suppose you have found a 10-minute video, but you only want to show the last 30 seconds of it. You would need to wait until the whole video stream is available, which can take several minutes, and then try to jump forward to where you want to start, which can be rather awkward.

Each of these problems can cost a lot of useful lecture time and produce a negative, distracting effect that can far outweigh the benefits that might be obtained by showing the video in the first place.

**3. Downloading and editing the video**
The most optimum approach would be to be able to download the online video to a local file, either obtained directly from the video's creator, or downloaded from a video archiving web site (e.g., *YouTube*). This would solve most of the problems related to showing the video using a web-interface, particular advantages would be:

- **Improved reliability:** In addition to being able to edit/crop the video (see below), the main advantage to download over online streaming video would be reliability. A video file stored on the hard-drive of the lecture theatre computer is guaranteed to play without the stutters and extended download breaks typical of streaming video;

- **Editing and Cropping:** In order to avoid class distraction, it would be ideal to be able to crop the video back to the bare minimum needed to get the point across. Since it is not possible to crop or edit online streaming video, this can only be done effectively with a downloaded local file. It would also be possible to couple short sections of video together to compare and contrast, etc., which would be very useful.

- **Optimal display:** Having a local file would also allow you to incorporate the movie directly into *PowerPoint* or other presentation software. Not only does this allow you to present the video more seamlessly (i.e., without having to break from presentation mode, find a browser window, etc.), but it also allows you to optimise the size of the displayed video on the screen such that its large enough to see without being so large that the movie looks grainy or pixelated.

- **Portability:** With a downloaded file, it is possible to show video content in lecture theatres that are either not equipped with internet or that don't have a fast internet connection.

**Legal issues revisited**

The three possible routes to presenting online video in lectures have very different legal implications. In each case, I will assume that the video in question does not violate copyright – an original work that either falls under the auspices of the Part VA/VB licence (or similar arrangements) or preferably has the consent of the copyright holder.

In the first case of simply presenting a hyperlink for students to follow, there are fewer issues and thus this is the safest option albeit the least effective. The use of the web sites involved by the students is considered personal use and would fall within the terms of use for almost all online video web sites. The second option depends largely on the website that you are using. For example, using a web site such as *YouTube*, this approach violates the terms of use because it involves public display, whereas with *GoogleVideo* there is no problem. In principle there should be no major copyright issues here, assuming the content meets the requirements already outlined above.

The same cannot be said for the third possibility, which would not only violate the terms of use for just about any online video web site, but it would also likely violate copyright unless the permission of the copyright owner was given to crop or edit the file and thereby create a derivative work, as well as to reproduce and publically display the resulting video. In the case where permission for unrestricted use of the video was granted, one viable option would be to obtain the video file directly from the copyright owner, thus removing the need to violate the terms of use for some online video web site when sourcing the video file. Note that this doesn't necessarily eliminate the benefit of the online video web site entirely, instead it simply shifts its role from being the provider of the video 'direct to student' or 'direct to class' to providing a means for educators to locate useful video content and identify its owner so that the content can be used for teaching purposes.

## Some motivations for using online video in lectures

In a physics course there are a number of good motivations for adding videos to a lecture. To conclude this article, I will present some of these potential motivations, in each case illustrated by some examples of videos posted to *YouTube* to give some inspiration for the potential of online video as an educational tool. These examples have vastly different origins, ranging from professionally produced videos and simulations through to amateur film clips made by children playing with their parents' digital video camera, demonstrating the diversity of content that can be useful from an educational perspective. In no particular order, some potential motivations include:

**Impractical demonstrations**

As discussed earlier, live demonstrations become a tougher prospect in higher year courses as the physics, equipment and techniques involved become more complex, making videos a good substitute. An excellent example is the demonstration of the properties of superfluid helium. Helium is a gas at room temperature, but when cooled below 2.2 Kelvin, it becomes a superfluid, a liquid that can flow without any resistance, giving it remarkable properties such as the ability to flow up walls against gravity, and through tiny holes and porous materials that other liquids cannot get through. Superfluidity is a quantum mechanical effect taught in many quantum physics courses worldwide, but one that it is almost never demonstrated live. This is in part because it is expensive and very difficult, but also because it requires a transparent glass container, so that the liquid can be seen, which has a bad habit of spontaneously exploding in a shower of glass and ultra-cold helium – certainly not great from a class safety perspective.

Video footage of the superfluid properties of helium is quite rare, and thus is not commonly shown in lectures. As a student, the best I ever saw were some small black and white photographs in a textbook, something that hardly conveys the fascinating nature of this liquid or how hard those properties are to observe. However, footage of superfluid liquid helium (YouTube 2007a) can now be found on *YouTube*, allowing today's students to actually see something that few of their lecturers have seen themselves.

Such demonstrations are important, even if they can't be done live, for two reasons. First they keep the science, which can sometimes become quite deep and technical, real and tangible by giving the students something that they can visualise. And second they show that there's more to physics than simple experiments with weights, strings, etc.; there are actually some very difficult, challenging and more interesting experiments to be done.

**Controlling time**

Another nice aspect of video is that you can do the impossible act of controlling time. You can speed it up; a nice example is time-lapse video of cloud formation (YouTube 2006a) demonstrating adiabatic cooling and convection in the atmosphere. You can also slow time down, for example, the classic bullet-strikes-object footage (YouTube 2007b). And you can even make time go backwards if you want (YouTube 2007c), for example, showing a firecracker in reverse, something that is really useful for demonstrating how the second law of thermodynamics, which states that entropy (i.e., disorder) always increases for irreversible processes, and thus sets an 'arrow' for time that defines forward (processes where entropy increases look normal) and backward (processes where entropy decreases, which look very strange).

The latter example is a good demonstration for why downloading and cropping video would be highly advantageous. To show the whole video would consume 10 minutes of the lecture and seriously bore the students, and to try to 'fast-forward' in the web interface would be rather impractical. This video is also a good demonstration that amateur footage should not be overlooked in searching for videos to show in physics lectures, in fact, some of the best examples of potential videos to use in lectures are amateur footage placed on *YouTube* for curiosity's sake.

**Visualisations**
*YouTube* is not just about video footage, there are a very large number of simulations and visualisations contributed to the public domain by various scientists and research groups. Examples range from how black holes act as gravitational lenses (YouTube 2007d) to the molecular machinery that replicates DNA (YouTube 2006b) to the classic double-slit experiment in quantum mechanics (YouTube 2007e).

**Question Setters**
Another potential use for *YouTube* videos is to set a question that introduces a new physics concept needing to be taught. A nice example is a phenomenon known as 'singing' in frozen lakes (YouTube 2006c), which is related to the Clausius-Clapeyron equation that describes phase transitions, such as that from water to ice. This is a fantastic home video (one of many on this phenomenon) that works well because the person who filmed it put the camera down on the ice, where the microphone picks up the sound very well. The question in this video is where these strange pinging and beeping noises come from. The answer is that water expands when it freezes, unlike most other substances, which contract. The noises are stress building up and releasing as various frozen sections push against one another. Beyond its possible use to set a question that motivates a more technical discussion, this video would have three other potential benefits. First it can help to make the concept more memorable, second it can demonstrate that even quite technical equations such as the Clausius-Clapeyron equation can be used to explain everyday things in nature such as freezing water, and finally, it can help stop a lecture from becoming just an hour of mathematics and technical discussion.

**Back-up for live demonstrations**
Live demonstrations in lectures don't always succeed. Hence another obvious potential application for *YouTube* videos is as a 'fallback' for live demonstrations that have a higher than usual probability of failing (e.g., electrostatics demonstrations on a humid summer afternoon). Video is no substitute for doing the demonstration itself, but on those occasions where Murphy's law gets you, it is good to have a fall-back to make sure you still manage to get your point across effectively.

**Humour and Interest**
Last but not least, *YouTube* videos can add some spice to your lectures and remind the students that what they are learning is fun. For example, in teaching engines in thermal physics, one possibility would be to show the singing Formula 1 car (YouTube 2006d), particularly to remind the engineering students in the course the fun that might lay ahead of them. Or in discussing probability and quantum information theory in quantum physics to show the Ricoh 'Intelligent models' commercial (YouTube 2007f) – the contents of which was allegedly plagiarised from the lectures of a MIT physics professor (Tadros 2007), perhaps an opportunity for the advertising world to repay their debt to the world of quantum physics!

Hopefully these few examples will convince you of the potential for online video as an educational tool, and that somewhere in the more than 59 million videos on *YouTube*, there is a video that might be of use for your course.

**Final thoughts – The need for a new era of Copyright**

> Over-regulation stifles creativity. It smothers innovation. It gives dinosaurs a veto over the future. It wastes the extraordinary opportunity for a democratic creativity that digital technology enables.
> (Lessig 2004; p.199)

As pointed out earlier, copyright aims to provide a balance between rewarding the creators of intellectual material while at the same time allowing reasonable and useful access to that material by consumers. This is a difficult balance to strike – too lenient and you remove the incentive that drives creativity, while too strict and you remove the material that fuels creativity. The key to establishing a balance in any system over time is the flexibility to adjust that balance where appropriate.

The same could be said for copyright law – as new technologies and circumstances evolve, the law needs to change to re-establish an appropriate balance. This has happened in the past, a classic example is the introduction of the Video Cassette Recorder (VCR) in the mid-1970s (von Lohmann 2005). Large media companies sought to have the VCR outlawed as a tool of piracy under the copyright laws that existed at the time. However, in the interests of fair use of content, the US Supreme Court instead amended the copyright legislation to allow the home taping of television shows for later viewing. This not only led to the widespread use of the VCR, which served as the foundation for modern video technologies such as DVD and online video, but also allowed the creation of new associated industries such as video sales and rental. Combined, these new opportunities provided massive rewards to the original complainants. Rewards they would otherwise not have gained had their lawsuit succeeded and that probably far outweigh the losses to piracy that motivated their lawsuit in the first place. Clearly this change to copyright was a wise decision on the part of the court at the time, one that has produced widespread benefit.

It is widely acknowledged that we are at a similar watershed in copyright law with the arrival of new digital media technologies and the information-sharing opportunities that have followed from the development of the Internet (Lessig 2004; von Lohmann 2005; O'Brien and Fitzgerald 2006; Atwood 2007; McCullagh 2007). In adapting copyright law to these new technologies, it is essential that we don't take the easy route, towards more draconian regulations that simply restrict free use of material more than ever before (Philipson 2008), out of an irrational fear of short term losses. We should instead take the somewhat harder route of creating new laws that provide greater freedom, because if history is anything to go by, they will bring far greater long-term benefits for all. There are many proposed mechanisms for achieving the latter option, and interestingly, in some cases it is the media companies themselves that are pioneering some of these ideas, such as open-access creative archives (British Broadcasting Corporation 2003).

One excellent idea is known as Creative Commons Licencing, which allows a creator to specify the level of protection that they wish to apply to their work, on a range that extends from full copyright (i.e., all rights reserved) through to public domain (i.e., no rights reserved) and a multitude of options in between. Creative commons licencing gives a creator incredible flexibility over how their work is used and protected by allowing them to choose some combination of four possible conditions to apply to their work, these are (Wikipedia 2008b):

- **Attribution:** A user may copy, distribute, display/perform the work and make derivative works providing they give the author due credit;
- **Non-commercial:** A user may copy, distribute, display/perform the work and make derivative works providing it isn't used for commercial purposes;
- **No derivative works:** A user may copy, distribute, display/perform only verbatim copies of the work and no derivative works can be made; and
- **Share-Alike:** A user may copy, distribute, display/perform the work and make derivative works providing any distributed works, verbatim or derivative, are distributed under a licence identical to that of the original work.

With an appropriate combination of these conditions it is easy to protect content such that it can be used 'not-for-profit' for educational or community purposes whilst at the same time not be unfairly exploited for financial gain.

In light of this, one potential route to achieving an online video archive for educational purposes would be to make a small modification to the video submission process for web sites such as *YouTube* that allows a user submitting material the option to declare ownership of the material (and in the case of derivative works made under creative commons licence, an attribution of the contributing works), and specify some requested level of protection under creative commons licence, be that full rights under existing copyright law if they so desire. In the latter case, users should also be able to specify if they are open to requests for permission to use the video or not. This information would be posted in the submission details for the video on the web site, enabling educators and other potential users to quickly identify the copyright requirements for a video and its potential for use. To some extent this is already occurring (YouTube 2006e; YouTube 2007g; YouTube 2007h; YouTube 2007i), but it would be more ideal if *YouTube* formally endorsed and encouraged this practice and provided new routes to facilitating its uptake (e.g., tickboxes in the file submission process, creative commons specific searches, etc.). Some simple changes like these would significantly streamline the process of finding digital videos suitable for use in lectures and be an important first step in realising the true potential of web sites such as *YouTube* as educational video archives.

## Acknowledgements

I'd like to thank Rosanne Quinnell, Michelle Kofod, Richard Newbury, Theodore Martin, Merlinde Kay and Debbie Gibson for helpful discussions and critical reading of this article. I'd also like to thank the referees for some very helpful suggestions for improving the article.